\documentclass[%
notitlepage,%
onecolumn,%
oneside,%
floats,%
aps,%
prd,%
nobibnotes,%
nofootinbib,%
amsmath,%
amssymb,%
amsfonts,%
amscd,%
superscriptaddress,%
eqsecnum%
]{revtex4-1}

\usepackage[utf8]{inputenc}
\usepackage{graphicx, array, dcolumn}
\usepackage[paperwidth=210mm, paperheight=297mm, centering,
hmargin=2.0cm, vmargin=2.4cm]{geometry}
\usepackage{ulem}

\begin{document}

\title{Evolution of thick domain walls in de Sitter universe}

\author{A.D.\,Dolgov}
\email{dolgov@fe.infn.it}
\affiliation{Novosibirsk State University, Novosibirsk, 630090, Russia}
\affiliation{Institute for Theoretical and Experimental Physics, Moscow, 117218, Russia}
\affiliation{Dipartimento di Fisica, Universit\`a degli Studi di Ferrara, I-44100 Ferrara, Italy}

\author{S.I.\,Godunov}
\email{sgodunov@itep.ru}
\affiliation{Novosibirsk State University, Novosibirsk, 630090, Russia}
\affiliation{Institute for Theoretical and Experimental Physics, Moscow, 117218, Russia}

\author{A.S.\,Rudenko}
\email{a.s.rudenko@inp.nsk.su}
\affiliation{Novosibirsk State University, Novosibirsk, 630090, Russia}
\affiliation{Budker Institute of Nuclear Physics, Novosibirsk, 630090, Russia}

\begin{abstract}
  We consider thick domain walls in a de Sitter universe following paper by Basu and
  Vilenkin~\cite{BV}. However, we are interested not only in stationary solutions
  found in~\cite{BV}, but also investigate the general case of domain wall evolution with time.
  When the wall thickness parameter, $\delta_0$, is smaller than $H^{-1}/\sqrt{2}$,
  where $H $ is the Hubble parameter in de Sitter space-time, then the stationary
  solutions exist, and initial field configurations tend with time to the stationary ones.
  However, there are no stationary solutions for $\delta_0 \geq H^{-1}/\sqrt{2}$.
  We have calculated numerically the rate of the wall expansion in this case and have found that
  the width of the wall grows exponentially fast for $\delta_0 \gg H^{-1}$.
  An explanation for the critical value $\delta_{0c} = H^{-1}/\sqrt{2}$ is also proposed.
 \end{abstract}

\maketitle

\section{Introduction}

As is well known, domain walls could be created in the universe
if/when a discrete symmetry is spontaneously broken. An interesting
example of this kind was suggested in ref.~\cite{lee-p-spont} where
the idea of spontaneous $CP$ violation was put
forward. However, in the simplest version this mechanism encounters
serious cosmological problems because even a single domain wall inside
the present day cosmological horizon would strongly distort the
observed isotropy of CMB~\cite{ZKO}. To cure this cosmological
disaster a few mechanisms of wall destruction were
proposed~\cite{kuzmin1, kuzmin2, kuzmin3, kuzmin4}. In our recent paper~\cite{dgrt} we
explored the idea of spontaneous $CP$ violation to construct a
(nearly) baryo-symmetric cosmology which might be compatible with
observations. According to this scenario domains with opposite signs of
$CP$ violation appeared during inflation and survived at the stage of reheating
when the baryogenesis operated. Later the walls between these domains dissolved
and therefore the domain wall problem did not arise.
As a result, this model could lead to baryo-symmetric universe with
cosmologically large regions of matter and antimatter.
Some general features of such dynamical $CP$
violation are described in refs~\cite{ad-dyn-CP-1, ad-dyn-CP-2, AD-Varenna}.

For successful implementation of such cosmological model it is imperative that
the distance between the matter-antimatter domains is also cosmologically large. It could be
realized if the width of the domain wall which existed during baryogenesis was cosmologically large
(we understand by the domain wall the piece of space between two regions in the universe,
where the field has not yet relaxed to its equilibrium value,
though the double minimum in the potential has already disappeared).
The matter-antimatter domains should be
separated by at least several megaparsec in terms of the present day
scale to avoid excessive matter-antimatter annihilation.
On the other hand, the distance should not be too large, otherwise the scenario would
lead to too large angular fluctuations of CMB~\cite{CdRG}.
It means in particular, that domains with opposite $CP$
symmetry breaking must be created during inflationary stage, otherwise
both the size of matter-antimatter domains and the transition regions
between them would be too small. In contrast, baryogenesis must proceed
after inflation was over to avoid strong dilution of the baryon asymmetry.

The evolution of the domain walls in de Sitter space-time was considered by Basu and
Vilenkin~\cite{BV}. The authors argued that the width of the
domain wall is determined by the ratio
$C \equiv \lambda\eta^2/H^2>0$, where $H$ is the Hubble parameter,
which was assumed to be constant,
$\eta$ is the vacuum expectation value of the Higgs-like field which induced the
spontaneous symmetry breaking,
and $\lambda$ is the coupling constant in the double-well potential, see
\eqref{eq:lagrangian}. If $C \gg 2$, the width of the domain wall
would be close to its flat space-time value, $\delta_0=1/(\sqrt{\lambda}\eta)$,
which is microscopically small, because $\sqrt{\lambda}\eta$  is essentially the mass
of the Higgs-like boson. To create astronomically wide domain wall this boson
must be practically massless and thus it would generate long range forces most probably excluded
or strongly restricted by experiment.

On the other hand, if $C < 2$, it is not excluded that the width of the domain wall may be astronomically large.
In presented paper we show that this is indeed the case.
In ref.~\cite{BV} only the stationary problem was considered, when
the shape of the domain wall was a function of a single variable, $l=zH\exp(Ht)$, which is
the length interval in de Sitter space. In this case the equation of motion
is reduced to an ordinary differential equation which makes the problem much simpler
technically. In the paper~\cite{BV} it was found numerically that the stationary solution exists only if $C>2$.

In what follows we lift the assumption of the stationarity and
consider the general plane solution being a function of both
variables: the distance from the wall and time. It allows us to see how the solution
approaches the stationary one and, in particular, what
happens with initial configurations if $C\leq 2$, when the stationary solution does not exist.

Our paper is organized as follows.
In Section~\ref{sec:stationary} we reproduce the calculation of Basu
and Vilenkin~\cite{BV} and provide a simple explanation why the
critical point is at $C=2$. In Section~\ref{sec:evolution} we consider
the field evolution with respect to both time and coordinate. Finally,
in Section~\ref{sec:conclusions} we conclude.

\section{Stationary solutions}
\label{sec:stationary}

In spatially flat section of de Sitter universe the expansion rate, $H=\dot{a}/a$, is constant
and the scale factor evolves as $a(t)=\exp{Ht}$. The FLRW metric for
such universe has the form:
\begin{equation}
ds^2 = dt^2-e^{2Ht}\left(dx^2+dy^2+dz^2\right).
\end{equation}

Let us consider a model of real scalar field $\varphi$ with the Lagrangian
\begin{equation}
\mathcal{L} = \frac{1}{2}g^{\mu\nu}
\partial_\mu\varphi\, \partial_\nu\varphi -
\frac{\lambda}{2}\left(\varphi^2-\eta^2\right)^2.
\label{eq:lagrangian}
\end{equation}

The corresponding equation of motion is
\footnote{There is a misprint in corresponding formula (3) in ref.~\cite{BV}.}
\begin{equation} \label{eq_of_mot}
\frac{1}{\sqrt{-g}} \partial_\mu \left(\sqrt{-g} g^{\mu\nu}
\partial_\nu\varphi \right) = -2\lambda\varphi\left(\varphi^2-\eta^2\right).
\end{equation}

In flat space-time, $H=0$, and in one-dimensional static case, $\varphi=\varphi(z)$,
the equation takes the form
\begin{equation}
\frac{d^2\varphi}{dz^2}=2\lambda\varphi\left(\varphi^2-\eta^2\right)
\end{equation}
and has a kink-type solution, which describes a static infinite domain
wall. Without loss of generality we can assume that the wall is situated at
$z=0$ in $xy$-plane:
\begin{equation}
\varphi(z)=\eta\,\tanh{\frac{z}{\delta_0}},
\end{equation}
where $\delta_0=1/(\sqrt{\lambda}\eta)$ has the meaning of the wall
thickness (subscript $0$ indicates that $H=0$).

Now let us consider an expanding universe with constant $H>0$. In this
case, if one looks for stationary solution, it is reasonable to
suggest that the field $\varphi$ depends only on $za(t)=z\exp{Ht}$,
which is the proper distance from the wall. So, one can choose the
following ansatz for $\varphi$:
\begin{equation}
\varphi=\eta\cdot f(u), \hspace{5mm} \mathrm{where} \hspace{3mm}
u=Hze^{Ht},
\end{equation}
where $u$ and $f$ are dimensionless.

With this ansatz the second independent variable, i.e. time $t$, does
not enter into the equation of motion which takes the form:
\begin{equation} \label{BV_eq}
\left(1-u^2\right)f''-4uf'=-2Cf\left(1-f^2\right).
\end{equation}
Here prime means the derivative with respect to $u$. It is noteworthy
that all parameters of the problem are combined into a single positive constant
$C=1/(H\delta_0)^2=\lambda\eta^2/H^2>0$.

Since we are interested in kink-type solutions, the boundary
conditions should be
\begin{equation} \label{bound_cond}
f(0)=0, \hspace{5mm} f(\pm\infty)=\pm1.
\end{equation}

\begin{figure}[h]
\includegraphics[width=0.49\textwidth]{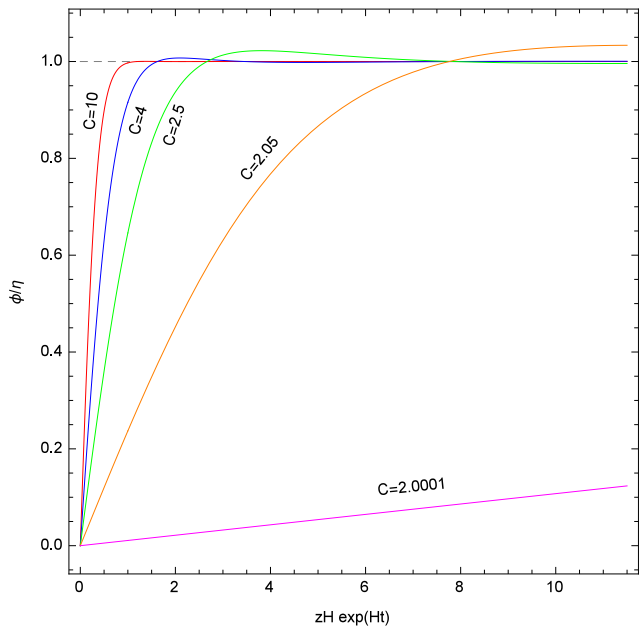}
\caption{\label{fig:BV_all} Stationary field configurations $f(u)$
for different values of parameter $C$.}
\end{figure}

Corresponding numerical solutions for different values of parameter
$C$ are shown in Fig.~\ref{fig:BV_all}. They are in good agreement
with those of ref.~\cite{BV}.
\footnote{There is a misprint in Fig. 1. of ref.~\cite{BV}, where the plot for $C=2.0001$ is presented,
but it is mistakenly labelled as $C=2.001$.}  We see that the
larger is $C$ the closer is the solution to the flat space-time one ($H=0$),
as is naturally expected.

As it is noticed in ref.~\cite{BV} the stationary solutions can be
found only for $C>2$, but no explanation of this observation is given
therein. Let us try to explain why the value $C=2$ is the special one.

From Eq.~(\ref{BV_eq}) and condition $f(0)=0$ it follows that
$f''(0)=0$. Using this condition and expanding $f(u)$ into Taylor series near
$u=0$, one obtains that for sufficiently small positive $\epsilon$ and for $f'(0)>0$:
\begin{eqnarray}
 f''(\epsilon)&<&0, \hspace{3mm} \mathrm{for} \hspace{2mm} C>2, \hspace{2mm} \mathrm{and}\\
 f''(\epsilon)&>&0, \hspace{3mm} \mathrm{for} \hspace{2mm} C\leq  2.
\end{eqnarray}

In other words, if the function $f(u)$ satisfies Eq.~(\ref{BV_eq}) and the
boundary conditions are $f(0)=0$ and $f'(0)>0$, then $f(u)$ is convex for $C>2$ and small
positive $u$, like ordinary kink-type solution is.  However, for
$C\leq 2$ the function $f(u)$ is concave, therefore it can not be the
kink-type one. So this is a simple explanation why in the case
$C\leq 2$ there are no stationary solutions of Eq.~(\ref{BV_eq}) with
boundary conditions~(\ref{bound_cond}). On the other hand, if we
allow for an arbitrary dependence of the solution on $z$ and $t$, it
exists for any $C$, but the case of $C\leq 2$ leads to the expanding kink
with rising width, as it is shown in the next section.

A more formal proof that there are no stable solutions for $C\leq 2$
can be found in Appendix \ref{sec:appendix_proof}.

\section{Evolution of domain walls beyond the stationary limit}
\label{sec:evolution}

As we have seen in the previous section, Eq.~(\ref{BV_eq}) allows
to find the field configurations, which describe stationary domain
walls in expanding universe. However, it is also interesting to see
how domain walls evolve from some initial states. Beyond the
stationary approximation we can find not only solution for $C>2$
but also for $ C\leq 2$, for which the stationary approximation does not
exist.

To this end one should solve the original equation of
motion~(\ref{eq_of_mot}) in the case when the field $\varphi$ is a
function of two independent variables, $z$ and $t$:
\begin{equation}
\frac{\partial^2\varphi}{\partial
t^2}+3H\frac{\partial\varphi}{\partial
t}-e^{-2Ht}\frac{\partial^2\varphi}{\partial z^2}
=-2\lambda\varphi\left(\varphi^2-\eta^2\right).
\label{eq-mot-phi}
\end{equation}

It is convenient to introduce dimensionless variables $\tau=Ht$,
$\zeta=Hz$ and function $f(\zeta,\tau)=\varphi(z,t)/\eta$. As a result
one obtains the equation
\begin{equation} \label{evolution}
\frac{\partial^2 f}{\partial \tau^2}+3\frac{\partial f}{\partial
\tau} -e^{-2\tau}\frac{\partial^2 f}{\partial \zeta^2} =2C
f\left(1-f^2\right),
\end{equation}
where $C=\lambda\eta^2/H^2=1/(H\delta_0)^2>0$ as it was above.

The boundary conditions for the kink-type solution should be
\begin{equation} \label{bound_cond1}
f(0,\tau)=0, \hspace{5mm} f(\pm\infty,\tau)=\pm1,
\end{equation}
and we choose the initial configuration as the domain wall with
''natural'' thickness $1/\sqrt{C}$ (with respect to dimensionless
coordinate $\zeta$) and zero time derivative:
\begin{equation} \label{in_cond1}
f(\zeta,0)=\tanh{\frac{z}{\delta_0}}=\tanh{\sqrt{C}\zeta},
\hspace{5mm} \frac{\partial f(\zeta,\tau)}{\partial
\tau}\biggl|_{\tau=0}=0.
\end{equation}

Of course, this is a toy model with the artificial initial
conditions. The realistic model should describe the evolution of
domain walls from the very beginning, including the process of wall
formation. This will be studied elsewhere.

We analyze solutions of Eq.~(\ref{evolution}) starting from
large values of parameter $C$ gradually moving to smaller ones.

\begin{figure}[t]
\begin{tabular}{c c}
\includegraphics[width=0.49\textwidth]{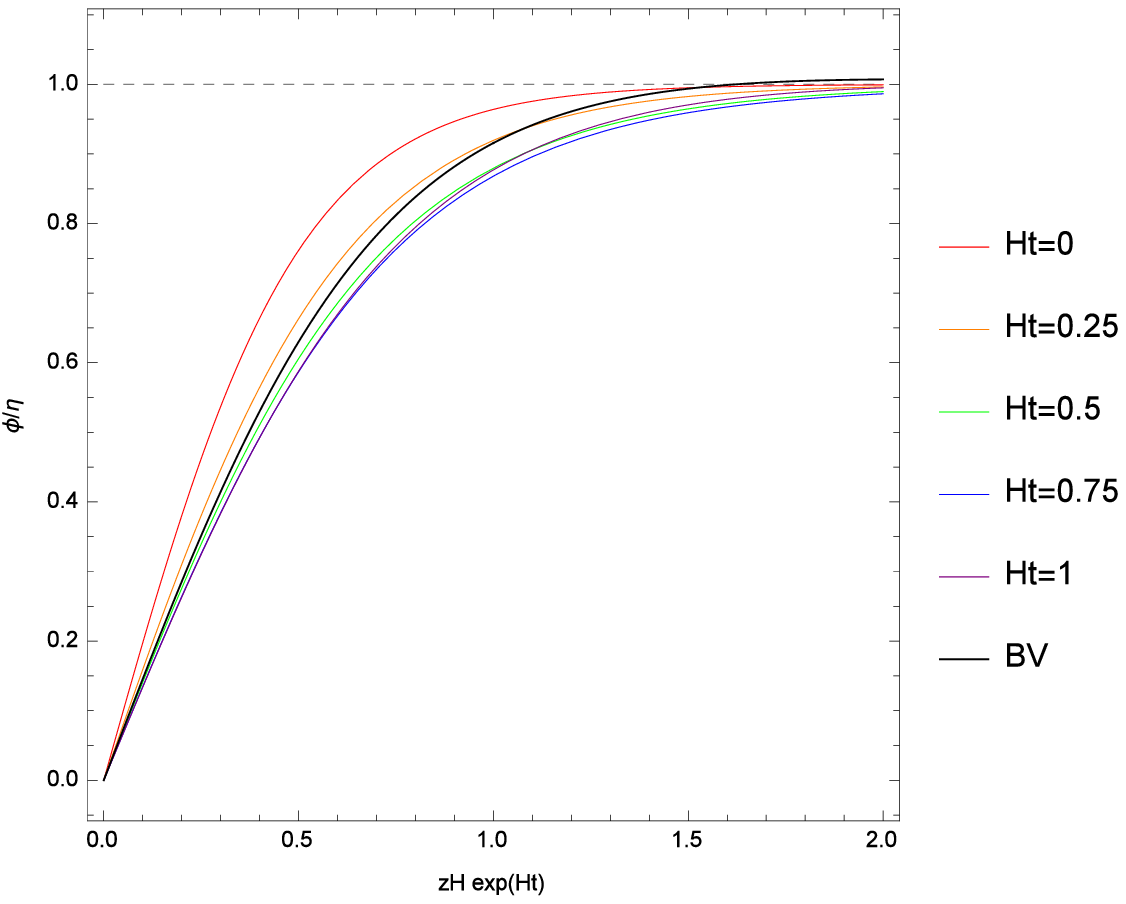} &
\includegraphics[width=0.49\textwidth]{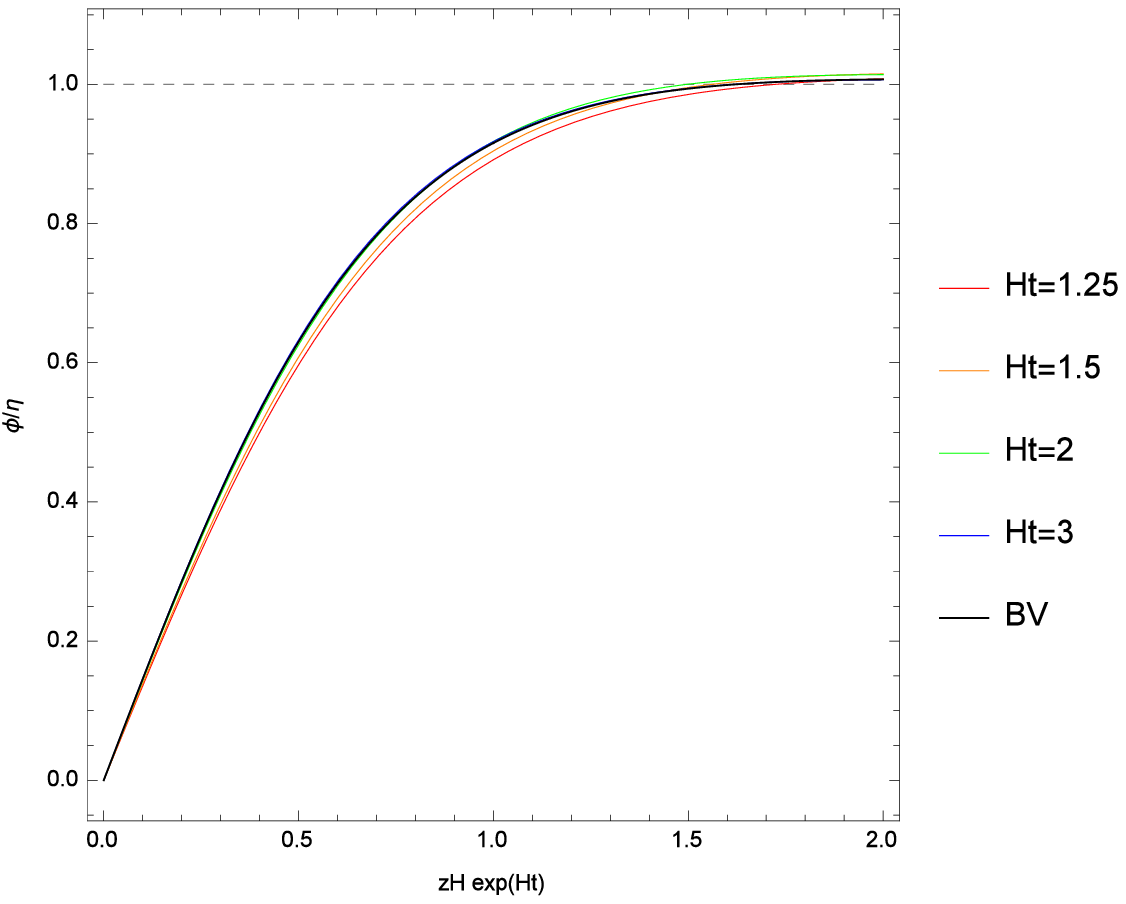}
\end{tabular}
\caption{\label{fig:C4} Evolution of domain wall for $C=4$. Black
curve corresponds to stationary solution.}
\end{figure}

The evolution of the domain wall for $C=4$ is depicted in
Fig.~\ref{fig:C4}. The stationary solution is shown there by black
curve (it is denoted by "BV" because of Basu and Vilenkin who found it
in ref.~\cite{BV}). One sees that the domain wall evolves from the
initial state~(\ref{in_cond1}) in somewhat non-trivial way. At the
very beginning the wall starts to broaden in terms of the proper
distance from the wall, $zH\exp{Ht}$, and at some moment it becomes
wider than the stationary solution.  However, afterwards the wall
broadening changes to contraction, here it occurs approximately at
$t\sim H^{-1}$ (but such "beautiful" value of $t$ seems to be
accidental). Finally, the wall comes to the stationary configuration
after several damped oscillations around it. The oscillating behavior
is expected because the field equation~(\ref{evolution}) is
the oscillator type equation with respect to $\tau$
and the value of the first time derivative is chosen arbitrarily.

\begin{figure}[th]
\begin{tabular}{c c}
\includegraphics[width=0.49\textwidth]{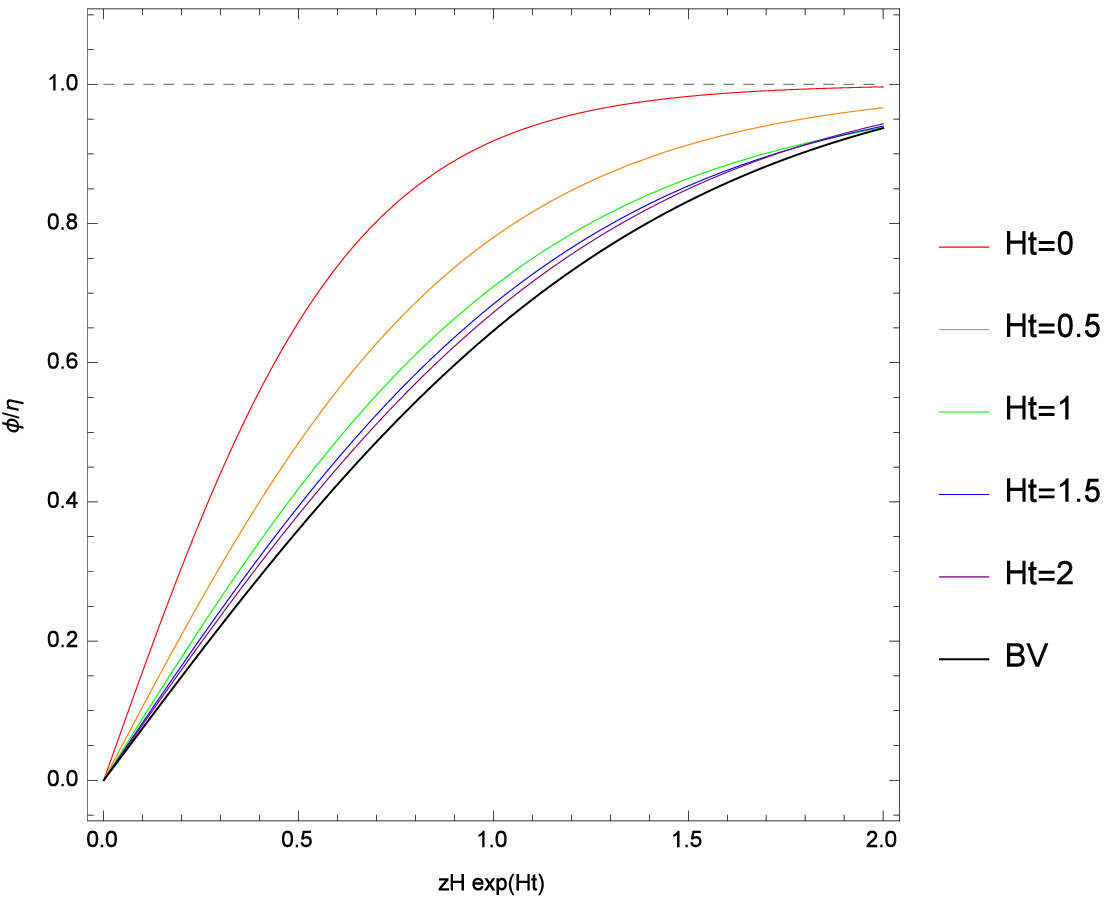} &
\includegraphics[width=0.49\textwidth]{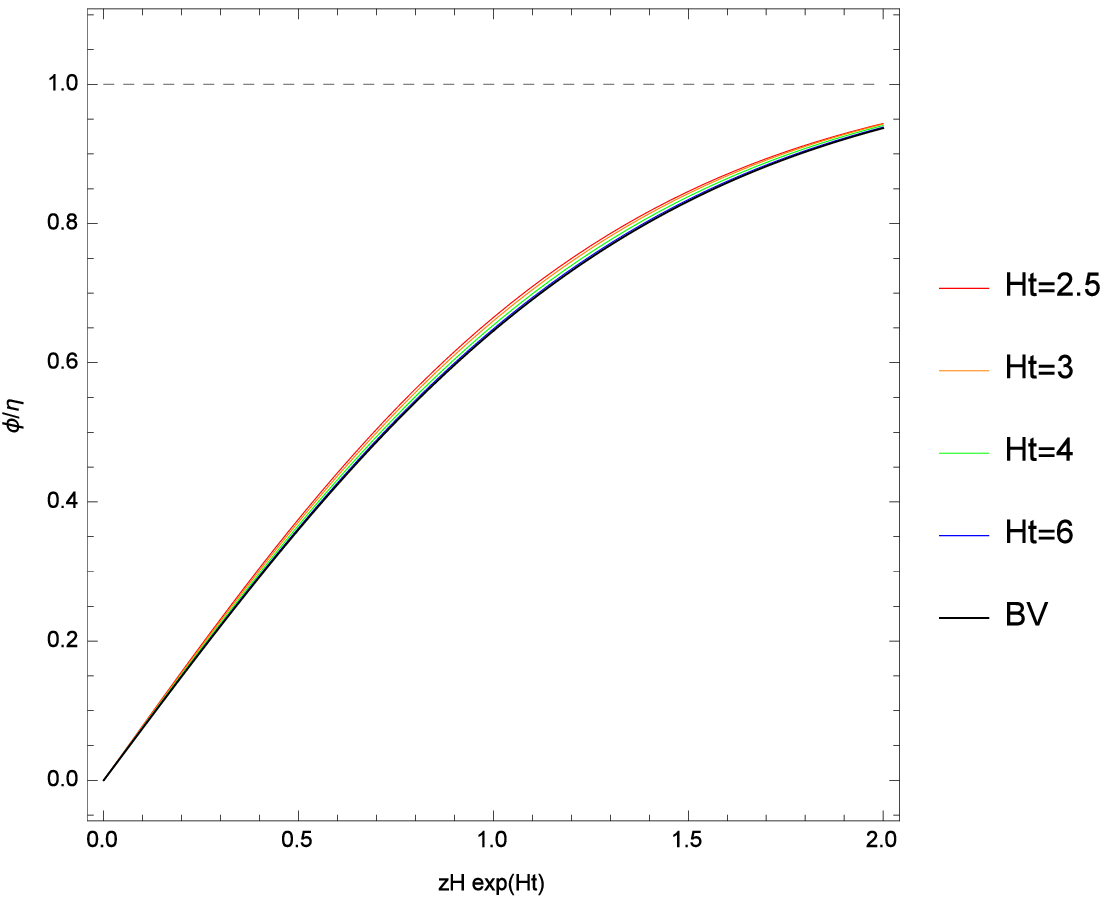}
\end{tabular}
\caption{\label{fig:C25} Evolution of domain wall for $C=2.5$. Initial
  configuration is $f(\zeta,0)=\tanh{\sqrt{C}\zeta}$.}
\end{figure}

Let us consider now smaller values of parameter $C$. When it is close to its
critical value $C=2$, the stationary domain wall is quite wide (see
Fig.~\ref{fig:BV_all}). Therefore, if the initial thickness of the
wall is $1/\sqrt{C}$~(\ref{in_cond1}), one should expect that the
non-stationary solution approaches to the stationary one only after quite
long time. The corresponding evolution of domain wall for $C=2.5$ is
depicted in Fig.~\ref{fig:C25}.

\begin{figure}[th]
\includegraphics[width=0.49\textwidth]{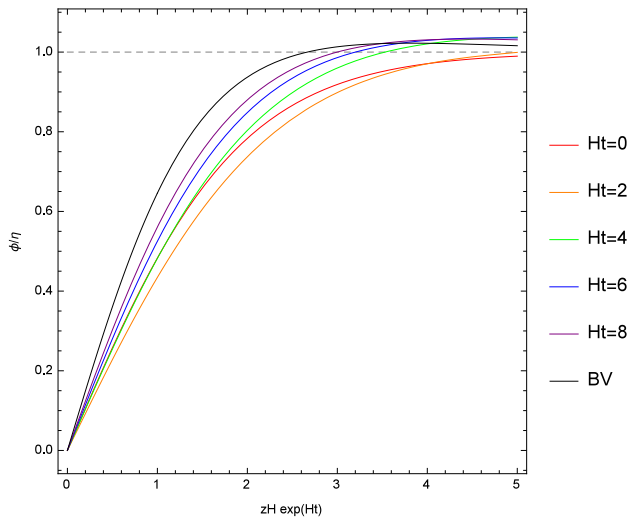}
\caption{\label{fig:C25_d} Evolution of domain wall for $C=2.5$.
  Initial configuration
   is $f(\zeta,0)=\tanh (\sqrt{C}\zeta/3)$.}
\end{figure}

It may be also interesting to choose an initial domain wall with the
thickness greater than that of the stationary solution. For such case
see Fig.~\ref{fig:C25_d}, where the evolution of domain wall with
initial thickness $3/\sqrt{C}$ for $C=2.5$ is presented.  The wall
also eventually comes to the stationary configuration.

\begin{figure}[th]
\begin{tabular}{c c}
\includegraphics[width=0.49\textwidth]{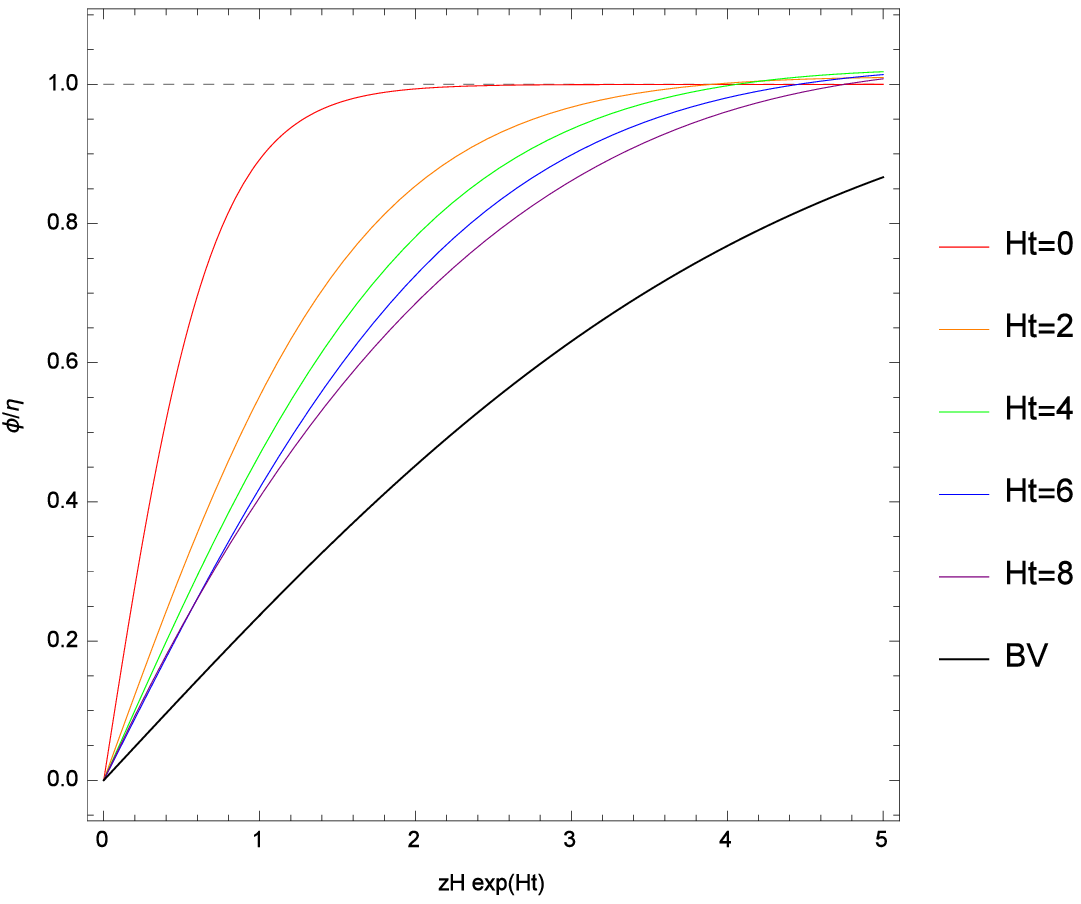} &
\includegraphics[width=0.49\textwidth]{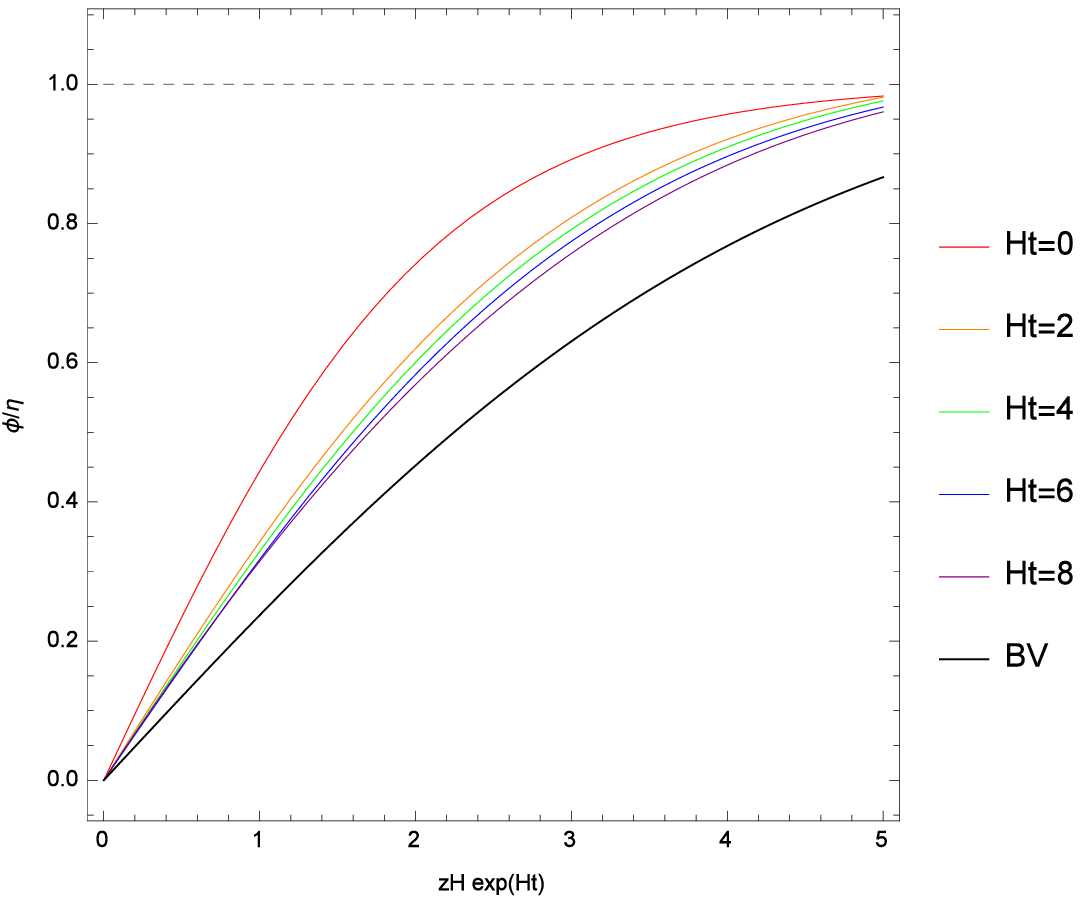}
\end{tabular}
\caption{\label{fig:C205} Evolution of domain wall for $C=2.05$.
  Initial configuration is $f(\zeta,0)=\tanh{(\sqrt{C}\zeta)}$ (left
  plot) and\\ $f(\zeta,0)=\tanh{(\sqrt{C}\zeta/4)}$ (right plot).}
\end{figure}

When $C$ is very close to 2, the solution of Eq.~(\ref{evolution})
converges to the stationary one very slowly.  One can see that in
Fig.~\ref{fig:C205} for $C=2.05$.

\begin{figure}[th]
\center
\begin{tabular}{c c}
\includegraphics[width=0.49\textwidth]{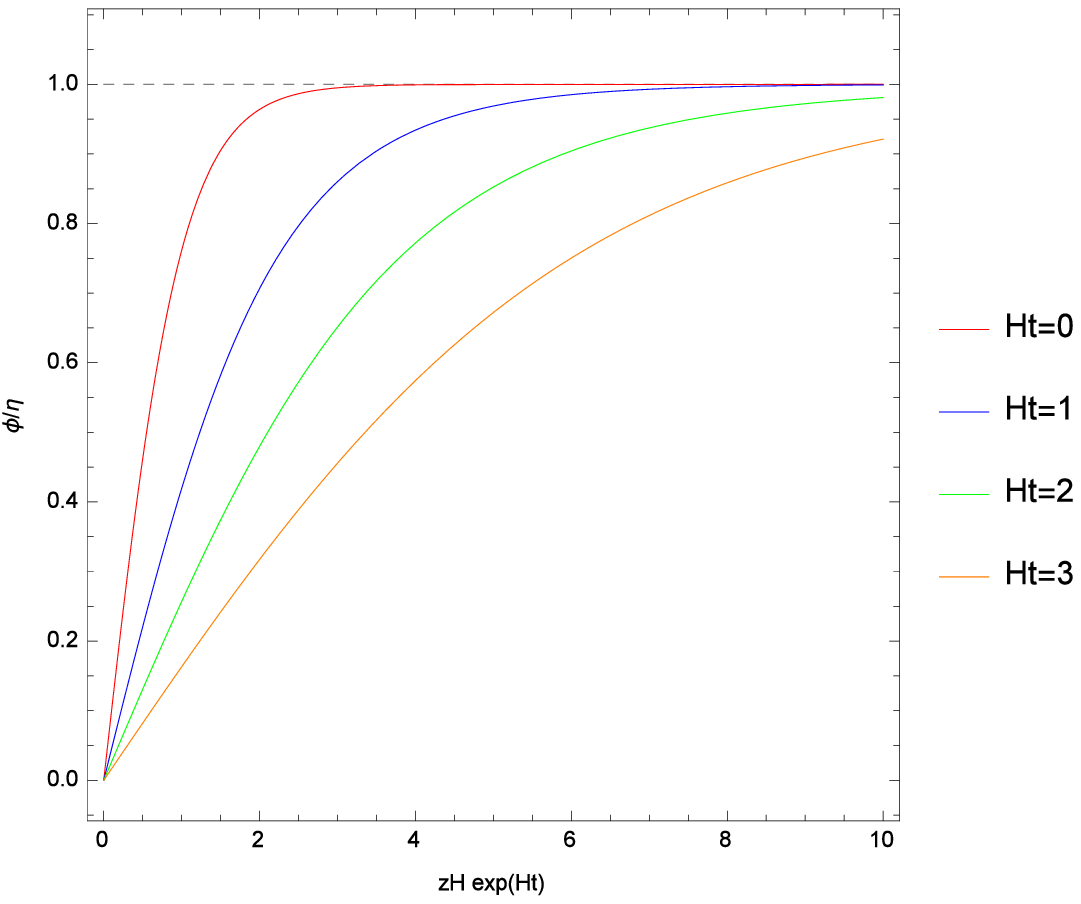} &
\includegraphics[width=0.49\textwidth]{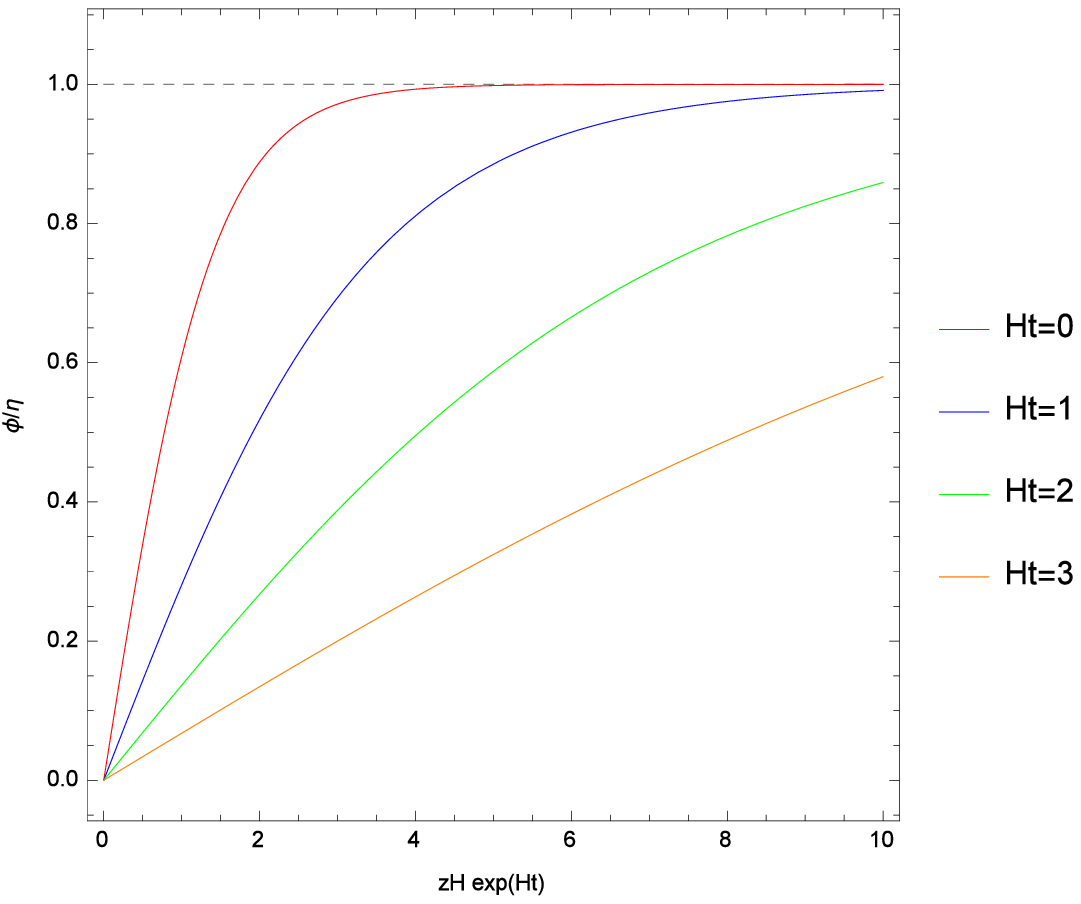}
\end{tabular}
\caption{\label{fig:C1_05} Evolution of domain wall for $C=1$ (left
  plot) and $C=0.5$ (right plot).}
\end{figure}

For $C\leq 2$ there are no stationary solutions at all. Evolution of
domain wall for such values of parameter $C$ is shown in
Fig.~\ref{fig:C1_05}. One can see that the domain wall thickness increases indeed.

In an earlier paper~\cite{dgrt} we assumed that
the size of the transition regions between baryon and antibaryon domains
(such regions formed in the place of the disappeared domain walls)
could be exponentially large, so the domains would be separated by
a few Mpc (in the present day scale).
Let us check now if this can be true and what is the proper magnitude of the parameters.
So we have to calculate how fast the domain wall thickness can increase in de Sitter universe.
In our model the wall is described at the initial moment by hyperbolic
tangent $\varphi/\eta=\tanh{(z/\delta)}$, where the wall thickness is
denoted by $\delta$. Although in the realistic course of the
evolution the wall is
not "pure" hyperbolic tangent, nevertheless one can use the same definition for
the thickness $\delta(t)$ as the value of the coordinate $z$ at the
position where the field $\phi$ reaches the value
$\phi/\eta=\tanh{1}\approx 0.76$.
\begin{figure}[ht]
\center
\begin{tabular}{c c}
\includegraphics[width=0.48\textwidth]{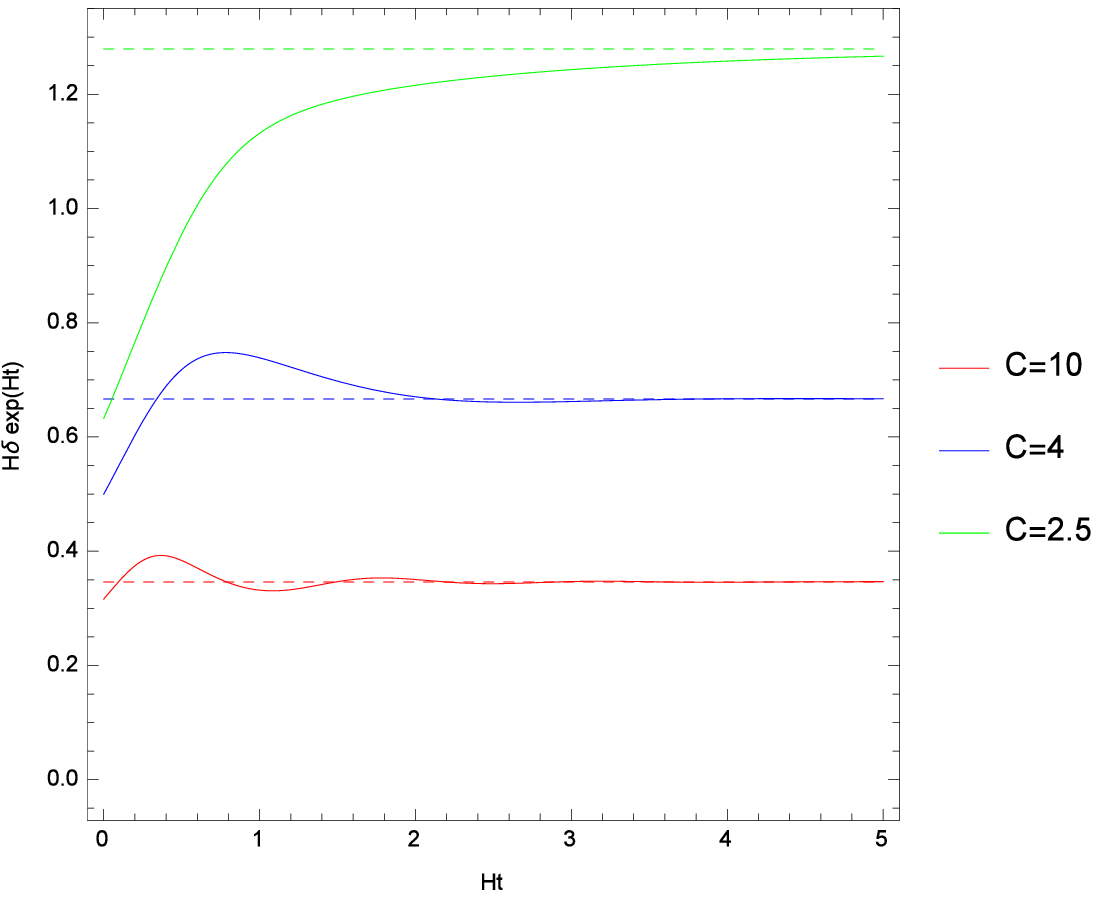} &
\includegraphics[width=0.50\textwidth]{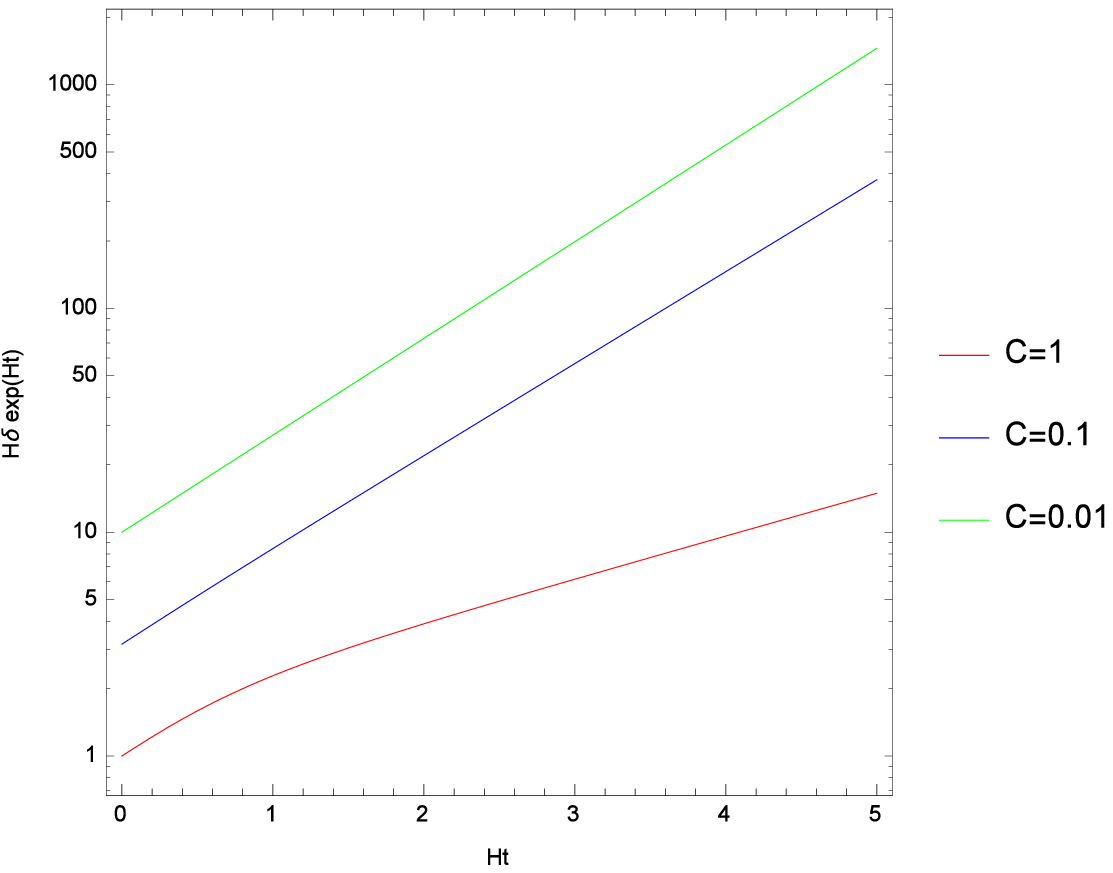}
\end{tabular}
\caption{\label{fig:thickness}
Time dependence of domain wall thickness for $C>2$ (left plot) and $C<2$ (right plot).
Dashed lines describe the stationary solutions.}
\end{figure}

In Fig.~\ref{fig:thickness} the time dependence of the physical width of
the wall in dimensionless units, $H\delta\exp{Ht}$,
for different values of $C$ is presented.
In the left plot ($C>2$) one can see that all the curves indeed tend
to constant values corresponding to stationary solutions (dashed
lines). Oscillatory behavior mentioned above is also apparent. Right
plot contains curves for $C<2$. Along the vertical axis
the wall thickness is shown in logarithmic scale
in order to compare the rate of the wall
expansion with the exponential cosmological one. It is clear that for $C=1$,
i.e. for not very small values of $C$, the wall thickness increases
slower than exponent. However, for smaller values of $C$, e.g. $C\lesssim 0.1$ the
rate of the wall expansion is the exponential one with a good accuracy.

The Hubble constant, $H$, plays the role of friction in this problem
and when it dominates over the potential term ($C\ll 1$), the field
configuration as a function of coordinates becomes almost
static. Therefore the width of the wall is growing nearly as the scale
factor $a(t)$, i.e. almost exponentially. This is why there are no
stable solutions for $C\ll 1$.

\section{Conclusions}
\label{sec:conclusions}

Time evolution of thick domain walls in a de Sitter universe is considered.
We have shown that for large values of parameter $C>2$
the initial kink configuration in a de Sitter background tends
to the stationary solution obtained by Basu and Vilenkin~\cite{BV}.
We also confirmed the BV result that the width of the stationary wall rises with decreasing value of $C$.

For $C<2$ the stationary solution does not exist and the width of the
wall infinitely grows with time. For $C \lesssim 0.1$ the rise is close to the
exponential one within the precision of our numerical
calculations. This result confirms the assertion made in
ref.~\cite{dgrt} that the transition region between matter and
antimatter domain might be cosmologically large.
This result is essential for application of spontaneous breaking of
symmetry between particles and antiparticles to realistic cosmology. In our version of this scenario~\cite{dgrt}
the walls between matter and antimatter domains dissolved and hence the huge energy density of domain
walls, which was a stumbling block of the traditional approach, did not destroy isotropy and homogeneity of
the universe.

Baryogenesis in our model proceeded after the double well potential returned to the
''normal'' potential with the single minimum at $\chi =0$ (here $\chi$ is field which made the wall).
Correspondingly the field $\chi$ started to move from the initial values
$\langle \chi \rangle = \pm \eta $ to zero, so the domain walls disappeared. However, if the classical field
$\chi$ did not completely relax down to zero prior to baryogenesis, it would induce $CP$-violation
of different signs at different domains which in turn led to an excess of matter or antimatter in this remnants
of the regions with non-zero $\chi$.

A necessary condition to make this model realistic is a sufficiently large distance between matter and antimatter
domains. Otherwise the annihilation on the boundaries would create too high gamma ray background,
according to ref.~\cite{CdRG}. The calculations presented above demonstrate that there is some range of the
parameter values, namely $C = \lambda \eta^2 /H^2 \lesssim 0.1$, for which the width of the domain wall
exponentially rises and so the distance between matter and antimatter could be large enough.

On the other hand, too wide domain walls lead to large universe regions devoid of baryons,
so it might distort the observed quasi-isotropy of CMB at the angular distance above the diffusion (Silk) damping scale.
However, this statement can be questioned, since the temperature contrast between baryon/antibaryon
and empty regions might push the domains apart diminishing the baryon-antibaryon diffusion towards each other
and this would allow for smaller separation between the domains below the diffusion damping scale.
The latter problem is under investigation now.

\vspace{5mm}

{\bf Note added.} After this paper was submitted to journal, we were informed about the
paper \cite{Voronov}, where the related problems were considered. We thank A.E. Kudryavtsev for this
reference.

\appendix
\section{Absence of stable solutions for $C\leq 2$}
\label{sec:appendix_proof}

Let us assume that equation (\ref{BV_eq})
\begin{equation}
  \label{eq:app1_f}
  \left(1-u^{2}\right)f''-4uf'+2Cf\left(1-f^{2}\right)=0
\end{equation}
with the boundary conditions (\ref{bound_cond})
\begin{equation}
f(0)=0, \hspace{5mm} f(+\infty)=1
\end{equation}
has a solution $f(u)$. We suppose that there is only one zero of
this solution: $f(0)=0$, i.e. $f(u)>0$ for $u>0$.
This assumption looks rather natural but we have not proven it.
If this is still true,
then due to the symmetry of eq.~(\ref{eq:app1_f}) with respect to "parity" transformation,
$u \to -u$, this zero must be at $u=0$. Accordingly the derivative $f'(0)$ must be positive.

Let us consider the corresponding linear equation
\begin{equation}
  \label{eq:app1_g}
  \left(1-u^{2}\right)g''-4ug'+2Cg=0
\end{equation}
with the boundary conditions
\begin{equation}
  \label{eq:app1_g_boundary}
  g(0)=0, \hspace{5mm} g'(0)=f'(0)>0.
\end{equation}
The solution of eq.~(\ref{eq:app1_g}) is close to $f(u)$ near $u=0$, when $f \ll 1$.

This equation can be easily solved analytically. For $C=2$ there is a special linear solution,
$g(u)=f'(0)u$, while for an arbitrary $C$ the solution is:
\begin{equation}
  \label{eq:app1_g_solution}
  g(u)=f'(0)\cdot u \cdot {_2F_1}\left(C_{5}^{-},C_{5}^{+};\frac{3}{2};u^{2}\right),
\end{equation}
where ${_2F_1}$ is the hypergeometric function and
\begin{equation}
  \label{eq:app1_C_n}
  C_{n}^{\pm}\equiv \frac{n\pm\sqrt{8C+9}}{4}.
\end{equation}

\begin{figure}[ht]
  \centering
  \includegraphics[width=0.50\textwidth]{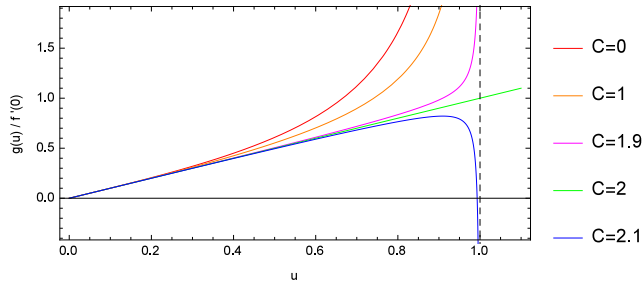}
  \caption{Solutions of equation \eqref{eq:app1_g} for different values of $C$.}
  \label{fig:app1_g_solutions}
\end{figure}

For $C<2$ one has $C_{5}^{\pm}>0$, therefore this hypergeometric function
goes to $+\infty$ when $u\to 1$.
The change in its behavior occurs when $C_{5}^{-}$ changes its sign, i.e. at $C=2$.
The solutions of eq.~(\ref{eq:app1_g}) for $C<2$, $C=2$ and $C=2.1$
with boundary conditions (\ref{eq:app1_g_boundary}) are presented in
Fig.~\ref{fig:app1_g_solutions}.

In what follows we will show that if $C<2$, then
$f(u)>g(u)$ in the interval $0<u<1$ and therefore the natural assumption
that the solution $f(u)$ is positive and bounded happens to be incorrect since if $C<2$, then
$g(u)\to +\infty$ when $u\to 1$ (see Fig.~\ref{fig:app1_g_solutions}).

The difference $\Delta(u)\equiv f(u)-g(u)$ satisfies the following
equation:
\begin{equation}
  \label{eq:app1_delta}
  \left(1-u^{2}\right)\Delta''-4u\Delta'+2C\Delta=h(u)\equiv 2Cf^{3}(u)
\end{equation}
with the boundary conditions
\begin{equation}
  \label{eq:app1_delta_boundary}
  \Delta(0)=0, \hspace{5mm}\Delta'(0)=0.
\end{equation}
Here we just want to verify the inequality $f(u)>g(u)$,
and therefore we are interested only in the {\it sign} of $\Delta(u)$.

Since the solution of the homogeneous equation vanishes due to the boundary
conditions \eqref{eq:app1_delta_boundary},
the solution of eq.~(\ref{eq:app1_delta}) is given by
\begin{equation}
  \label{eq:app1_delta_solution}
  \Delta\left(u\right) =
  \int\limits_{0}^{+\infty}h(v)G(u,v)dv,
\end{equation}
where $G(u,v)$ is the Green's function of eq.~(\ref{eq:app1_delta}). It satisfies:
\begin{equation}
  \label{eq:app1_delta_v}
  \left(1-u^{2}\right)G''-4uG'+2CG=\delta(u-v)
\end{equation}
with boundary conditions \eqref{eq:app1_delta_boundary}.
Here prime means derivative with respect to $u$ and $\delta(u-v)$ is the Dirac delta function.

The solution of \eqref{eq:app1_delta_v} can be found analytically (we
are interested in $G(u,v)$ only for $0<u<1$):
\begin{equation}
  \label{eq:app1_delta_v_solution}
  G(u,v)=\frac{3\Theta\left(u-v\right)}{\left(1-v^{2}\right)}
  \frac{uI\left(3;\frac{1}{2};v^{2}\right)I\left(5;\frac{3}{2};u^{2}\right)-vI\left(3;\frac{1}{2};u^{2}\right)I\left(5;\frac{3}{2};v^{2}\right)}
  {6Cv^{2}I\left(5;\frac{3}{2};v^{2}\right)I\left(7;\frac{3}{2};v^{2}\right)+I\left(3;\frac{1}{2};v^{2}\right)\left[3I\left(5;\frac{3}{2};v^{2}\right)+2(2-C)v^{2}I\left(9;\frac{5}{2};v^{2}\right)\right]},
\end{equation}
where $\Theta(u-v)$ is the Heaviside step function and we have introduced the following
notation:
\begin{equation}
  \label{eq:app1_I}
  I\left(n;c;u^{2}\right)\equiv {_2F_1}\left(C_{n}^{-},C_{n}^{+};c;u^{2}\right).
\end{equation}

\begin{figure}[ht]
  \centering
  \includegraphics[width=0.49\textwidth]{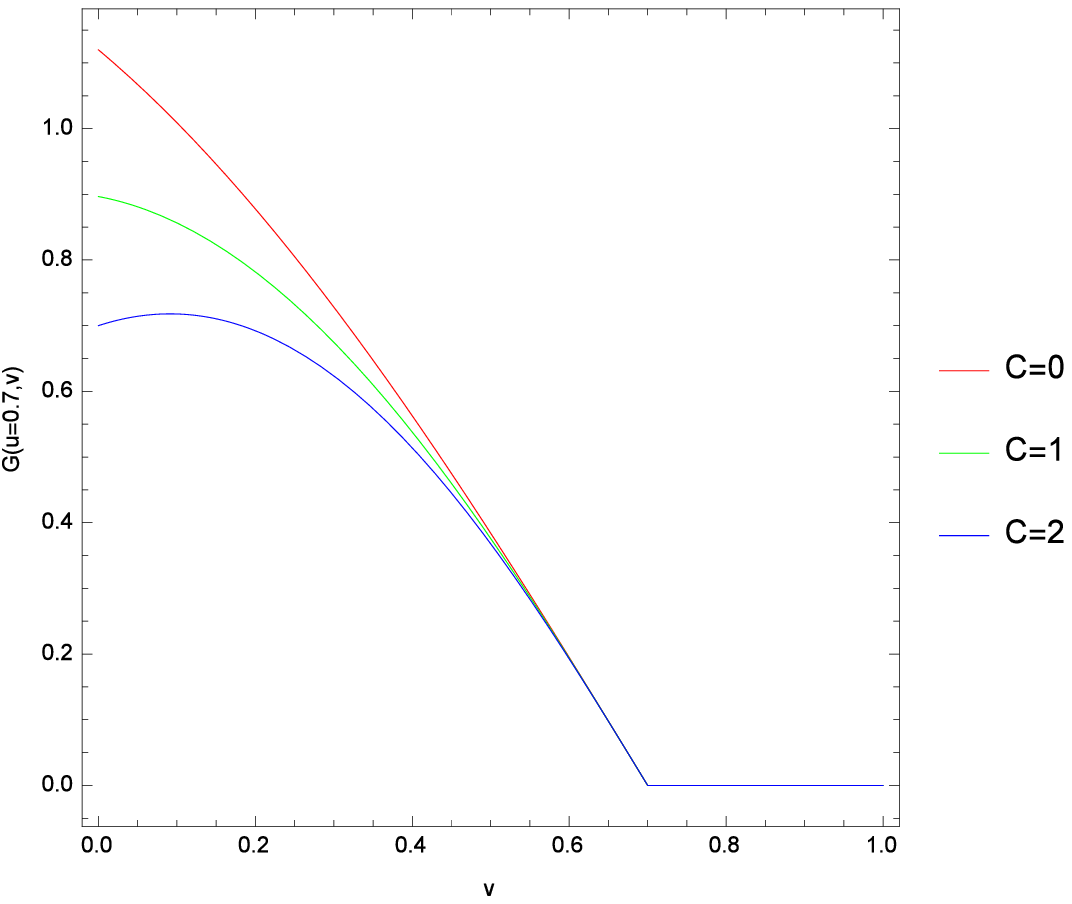}
  \includegraphics[width=0.49\textwidth]{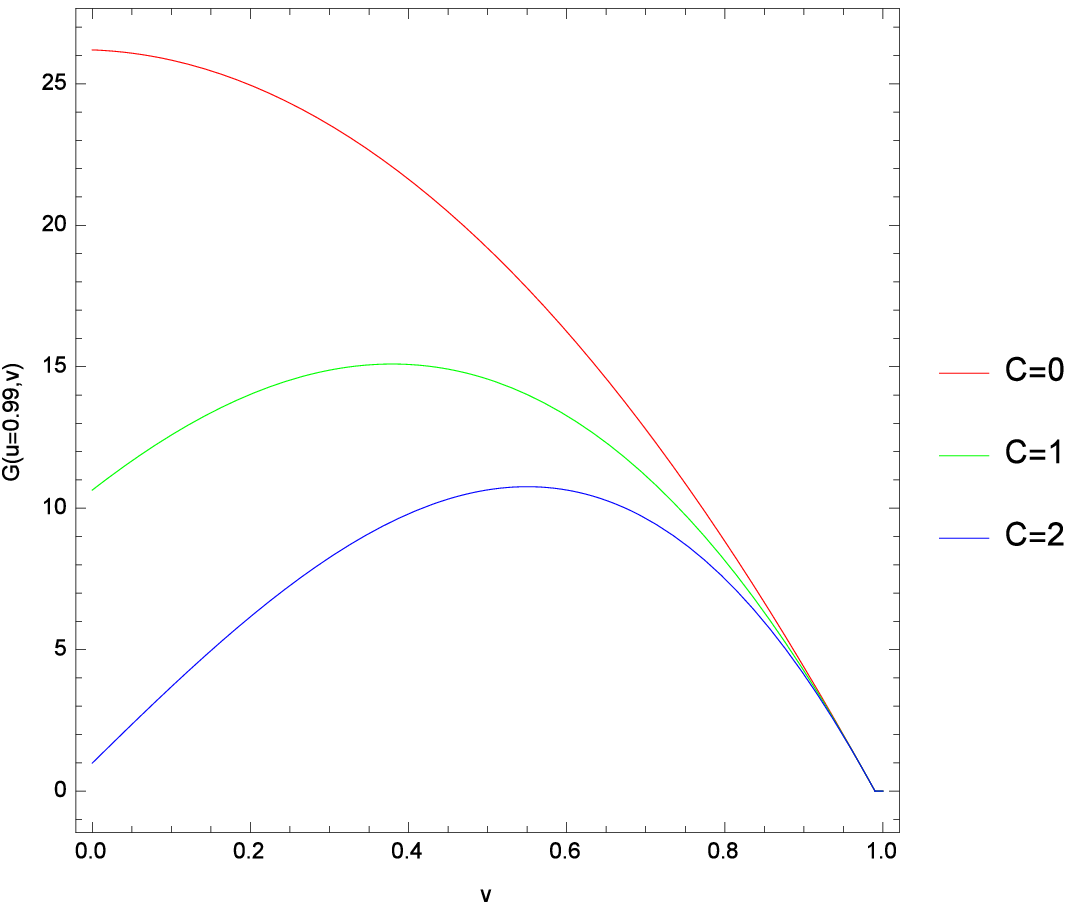}
  \caption{Green's function $G(u,v)$ for $u=0.7$ (left) and $u=0.99$ (right).}\label{fig:app1_G_v}
\end{figure}

Studying the Green's function one can find that it is non-negative,
$G(u,v)\geq 0$, for $0<u<1$ and $C<2$.
For the illustration of that see Fig.~\ref{fig:app1_G_v} where $G(u,v)$
as function of $v$ for a couple of different constant values of $u$ is presented.
Since $h(v)\geq 0$ according to our assumption, one sees from \eqref{eq:app1_delta_solution}
that $\Delta(u)\geq 0$ for $0<u<1$.
As it was mentioned above, it proves that there are no solutions of eq.~\eqref{eq:app1_f} for $C<2$.

\begin{figure}[ht]
  \centering
  \includegraphics[width=0.49\textwidth]{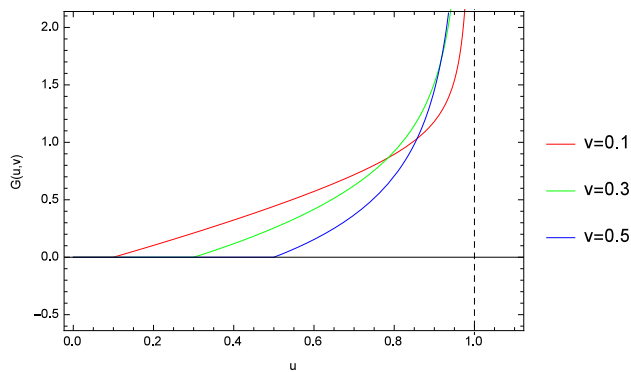}
  \caption{Green's function $G(u,v)$ for different values of $v$ in the case $C=2$.}
  \label{fig:app1_G_u}
\end{figure}

In the case of $C=2$ it is not enough to find that $f(u)>g(u)$ since there is no
singularity in $g(u)$ at $u=1$. But $G(u,v)$ itself has a positive
singularity at $u=1$ for any $v>0$ (see Fig.~\ref{fig:app1_G_u}).
Then it follows from \eqref{eq:app1_delta_solution} that $\Delta(u)$ goes to $+\infty$ at $u=1$
and therefore for $C=2$ there are no stable solutions of eq.~\eqref{eq:app1_f} as well.

\section*{Acknowledgements}
We acknowledge support of the Grant of President of Russian Federation
for the leading scientific Schools of Russian Federation,
NSh-9022-2016.2. SG is also supported under the grants RFBR
No.~16-32-60115, 14-02-00995, 16-02-00342, and by the Russian
Federation Government under the grant MK-4234.2015.2. In addition, SG
is grateful to Dynasty Foundation for support.

\end{document}